\documentclass[aps,prx,twocolumn,superscriptaddress,letterpaper,longbibliography]{revtex4-1}

\usepackage{amssymb}
\usepackage{amsmath}
\usepackage{bm}
\usepackage{graphicx}
\usepackage{color}
\usepackage{epsfig}
\usepackage{ulem}
\usepackage{siunitx}
\usepackage{float}
\usepackage{enumerate}

\begin{document}

\title{Amplification of quantum signals by the non-Hermitian skin effect}

\author{Qiang Wang}
\affiliation{Division of Physics and Applied Physics, School of Physical and Mathematical Sciences, Nanyang Technological University,
Singapore 637371, Singapore}

\author{Changyan Zhu}
\affiliation{Division of Physics and Applied Physics, School of Physical and Mathematical Sciences, Nanyang Technological University,
Singapore 637371, Singapore}

\author{You Wang}
\affiliation{Division of Physics and Applied Physics, School of Physical and Mathematical Sciences, Nanyang Technological University,
Singapore 637371, Singapore}

\author{Baile Zhang}
\email{blzhang@ntu.edu.sg}
\affiliation{Division of Physics and Applied Physics, School of Physical and Mathematical Sciences, Nanyang Technological University,
Singapore 637371, Singapore}
\affiliation{Centre for Disruptive Photonic Technologies, Nanyang Technological University, Singapore, 637371, Singapore}

\author{Y. D. Chong}
\email{yidong@ntu.edu.sg}
\affiliation{Division of Physics and Applied Physics, School of Physical and Mathematical Sciences, Nanyang Technological University,
Singapore 637371, Singapore}
\affiliation{Centre for Disruptive Photonic Technologies, Nanyang Technological University, Singapore, 637371, Singapore}

\begin{abstract}
  The non-Hermitian skin effect (NHSE) is a phenomenon whereby certain non-Hermitian lattice Hamiltonians, particularly those with nonreciprocal couplings, can host an extensive number of eigenmodes condensed to the boundary, called skin modes.  Although the NHSE has mostly been studied in the classical regime, we show that it can also manifest in quantum systems containing boson number nonconserving processes arising from uniform parametric driving.  We study lattices of coupled nonlinear resonators that can function as reciprocal quantum amplifiers.  A one-dimensional chain exhibiting NHSE can perform strong photon amplification, aided by the skin modes, that scales exponentially with the chain length and outperforms alternative lattice configurations that lack the NHSE.  We show also that two-dimensional nonlinear lattices can perform directional photon amplification between different lattice corners, due to the two-dimensional NHSE.
\end{abstract}

\maketitle
\section{Introduction}

Non-Hermitian dynamical systems, which do not obey energy conservation, have long been known to act in qualitatively different ways from Hermitian ones.  In photonics, non-Hermiticity can enter via the damping or amplification of electromagnetic waves, and leads to distinctive phenomena \cite{bender1998, mostafazadeh2002, feng2017, leykam2017, el2018, shen2018, ozdemir2019, yokomizo2019, kawabata2019, bergholtz2021} such as coherent perfect absorption \cite{Chong2010, baranov2017, wang2021coherent} and parity/time-reversal symmetric lasing \cite{feng2014, Parto2018, Zhu2021}.  Photonics has also been used to explore theoretically noteworthy aspects of non-Hermiticity, such as the embedding of exceptional points in band structures \cite{zhen2015spawning, chen2017exceptional, cerjan2019experimental}.  One non-Hermitian phenomenon that has acquired particular prominence in recent years is the non-Hermitian skin effect (NHSE) \cite{Hatano1996, Hatano1997, Yao2018}, which represents a breakdown of standard bulk-boundary correspondence principles \cite{Bansil2016}: a finite lattice with open boundary conditions (OBC) can host an extensive number of localized ``skin modes'', totally unlike the extended Bloch modes formed under periodic boundary conditions (PBC) \cite{Yao2018}.  In one-dimensional(1D) lattices, the NHSE can be related to a non-Hermitian topological band invariant, a nonzero point gap winding formed by the complex energy spectrum \cite{Gong2018, Lee2019Anatomy, Okuma2020, Zhang2020}. In higher dimensions, the relationship between bandstructures and the NHSE remains under investigation \cite{Lee2019, Borgnia2020, zhang2021}.  The NHSE has been realized in photonic systems \cite{weidemann2020, xiao2020, xiao2021}, as well as in classical electrical \cite{zou2021, helbig2020}, acoustic \cite{Zhang2021Observation, zhang2021acoustic}, and mechanical \cite{ghatak2020} metamaterials, though the application possibilities for this intriguing phenomenon are still unclear.  Both the original theoretical formulation of the NHSE and the above experimental demonstrations have been based on single-particle models, applicable to non-interacting quantum particles or linear classical waves.  Only recently have researchers begun to explore the NHSE beyond the single-particle regime, such as in interacting spin-less fermion chains \cite{luo2020skin}, driven-dissipative cavity arrays \cite{wanjura2020, Wanjura2021} and correlated boson systems \cite{lee2020many,zhang2021manybody, Xu2021, Yokomizo2021, Okuma2022}.

Bosonic systems governed by quadratic Hamiltonians have drawn attention as an unusual and interesting way to access non-Hermitian dynamics.  Such many-body Hamiltonians can be mapped to single-particle Hamiltonians via the Bogoliubov-de Gennes (BdG) transformation \cite{Rossignoli2005}, a well-known procedure used in studying Majorana fermions \cite{kitaev2001} and topological superconductors \cite{Qi2010}.  The transformed BdG Hamiltonian can be both non-Hermitian and nonreciprocal even if the underlying many-body Hamiltonian is Hermitian and reciprocal \cite{McDonald2018, Wang2019, Xu2021}, which is especially noteworthy for the NHSE since it often depends on the presence of nonreciprocal couplings \cite{Hatano1996, Hatano1997, Yao2018, Gong2018, Lee2019Anatomy, Okuma2020, Zhang2020, Lee2019, Borgnia2020, zhang2021}.  Previous authors have explored using the dynamics of BdG Hamiltonians for quantum amplification and related purposes \cite{Caves1982, Rossignoli2005, Clerk2010, Caves2012,McDonald2018, Xu2021}, including showing that BdG Hamiltonians can be made to exhibit distinct topological phases with unidirectional and/or amplified topological modes \cite{barnett2013, Engelhardt2015, Galilo2015, Engelhardt2016, peano2016, Bardyn2016, peano2016NC, Lieu2018, roy2021}. While most of these studies have been theoretical, quantum amplifiers have been successfully observed in simple quantum circuits \cite{Abdo2014, Sliwa2015, macklin2015}.

\begin{figure*}
\centering
\includegraphics[width=0.95\textwidth]{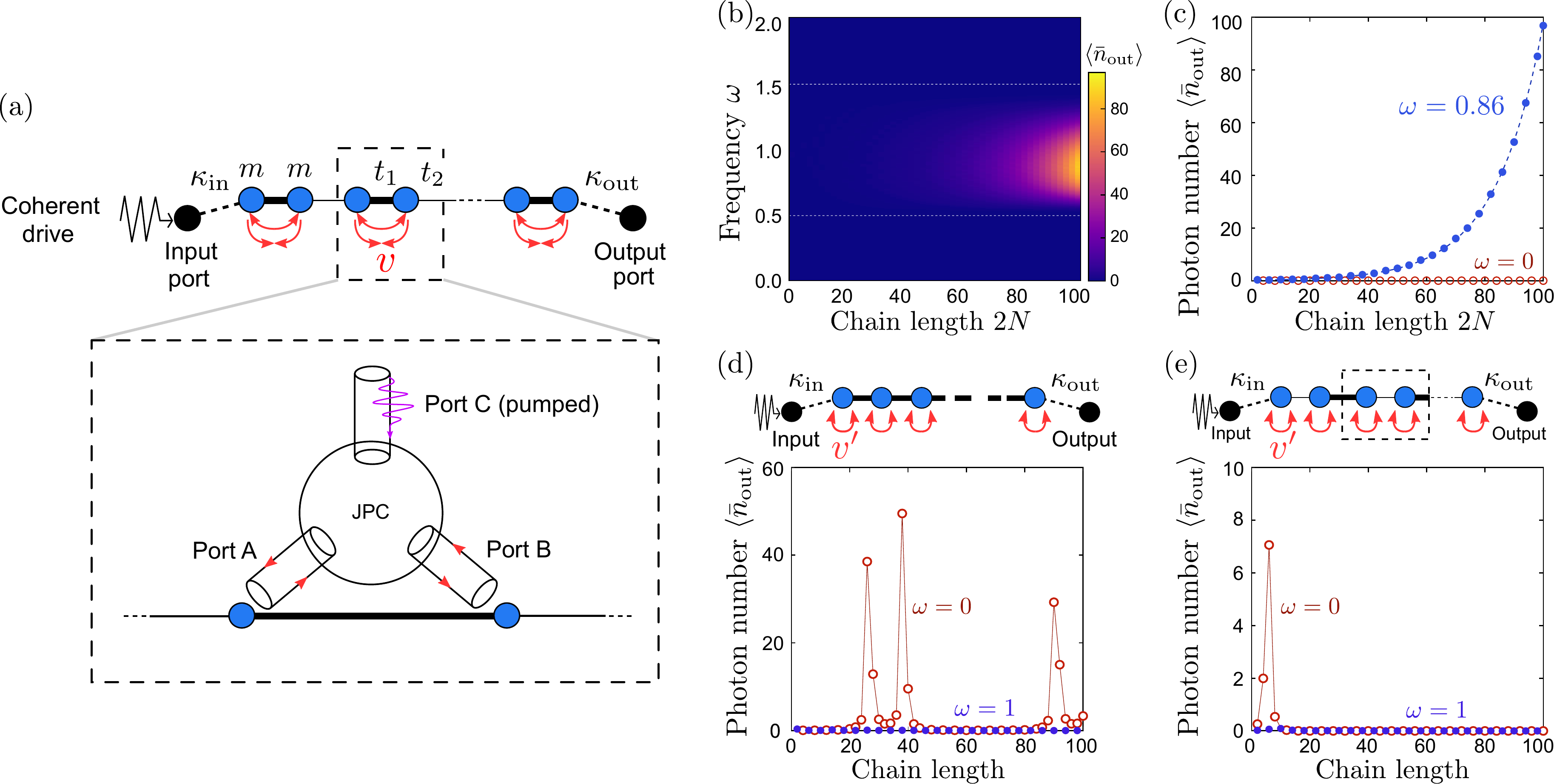}
\caption{(a) Top panel: schematic of a 1D chain with $2N$ sites (blue circles), two sites per unit cell, intracell/intercell hoppings $t_1$ and $t_2$ (thick/thin lines), and on-site mass $m$.  Each unit cell has two-boson creation/annihilation processes with coefficient $v$ (red arrows). Input and output ports (black circles) are coupled to the leftmost and rightmost sites with hoppings $\kappa_{\mathrm{in}}$ and $\kappa_{\mathrm{out}}$. A steady-state coherent drive, of unit amplitude and frequency $\omega$, is applied to the input. Bottom panel: a possible realization of the nonlocal nonlinearity using a three-port Josephson parametric convertor (JPC) with one port pumped.  (b) Heat map of the expected mean photon number at the output port, $\langle\bar{n}_{\mathrm{out}}\rangle$, versus driving frequency $\omega$ and chain length $2N$. (c) Plot of $\langle\bar{n}_{\mathrm{out}}\rangle$ versus chain length at the optimal operating frequency $\omega = 0.86$ (blue circles) and at zero detuning (red circles).  The lattice parameters in (b)--(c) are $m=0$, $t_1=1$, $t_2=0.5$, $\kappa_{\mathrm{in}}=0.7$, $\kappa_{\mathrm{out}}=0.1$, $v = 0.1$, and $\gamma = 0.02$. Dashes show the fitted exponential $\langle\bar{n}_{out}\rangle=0.22\exp\left(0.061 \cdot 2N\right)$ for the $\omega = 0.86$ case.  (d)--(e) Comparison to two alternative models that lack the NHSE: (d) a lattice with one site per unit cell and inter-site hoppings $t_1$, with on-site two-boson interaction; and (e) a dimer chain with flipped intracell and intercell hoppings of $t_2$ (thick lines) and $t_1$ (thin lines) respectively, and on-site two-boson interaction.  As shown in the lower panels, in both cases there is only significant photon amplification at certain lattice sizes, and only at $\omega \approx 0$ where the modes are unstable.  The parameters for (d)--(e) are $m=0$, $t_1=1$, $t_2=0.5$, $\kappa_{\mathrm{in}}=0.7$, $\kappa_{\mathrm{out}}=0.1$, $\gamma = 0.02$, and $v' = 0.05$.
}
\label{fig:f1}
\end{figure*}

In this paper, we show that the NHSE can manifest in lattices of resonators with nonlinear two-boson interactions, and that this provides a way to construct one- or higher-dimensional arrays of quantum amplifiers.  The lattices are described by bosonic quadratic Hamiltonians and can be implemented with microwave quantum circuits containing parametrically driven Josephson junctions \cite{makhlin2001, Anderson2016}, or other quantum optical platforms \cite{Ilchenko2004, guarino2007, luo2019}.  By designing a lattice so that the BdG Hamiltonian exhibits the NHSE, strong and robust photon amplification can be achieved without isolators.  With a conventional lattice design that does not support the NHSE, similar amplification cannot be achieved except via special fine-tuning.  Previously, McDonald \textit{et al.}~have found similar amplifying behavior in 1D lattices implementing the bosonic Kitaev-Majorana chain \cite{McDonald2018}.  Unlike that work, the models we present do not require nonreciprocal inter-resonator couplings, and exhibit reciprocal amplification under uniform parametric driving of all the resonators in the lattice.  The required two-boson nonlinearities can be implemented using a relatively simple configuration of three-port Josephson parametric convertors \cite{bergeal2010, Sliwa2015}.  We show that the NHSE-induced amplification occurs not only in 1D lattices, but also in two-dimensional (2D) lattices where the BdG Hamiltonian gives rise to a corner NHSE \cite{Lee2019, Zhang2021Observation, zou2021}; this can be used to achieve directional amplification between different corners of a 2D sample.

\section{One-dimensional model}

The top panel of Fig.~\ref{fig:f1}(a) depicts a 1D dimer chain, based on the Su-Schrieffer-Heeger (SSH) model \cite{Su1979}, with real intracell (intercell) hoppings $t_1$ ($t_2$) drawn as thick (thin) lines.  There is also an on-site mass $m$, and two-boson creation/annihilation processes indicated by red arrows.  The Hamiltonian is
\begin{align}
\begin{aligned}
  \mathcal{H} &=
  \sum_{n=1}^N \Bigg[ \left(t_1 a_{2n-1}^\dagger a_{2n}+t_2 a_{2n-1}^\dagger a_{2n-2}+ \mathrm{h.c.}\right)\\
    &\quad\qquad
    + \left( v\, a_{2n-1}^\dagger a_{2n}^\dagger+ \mathrm{h.c.}\right) \Bigg]
  + m \sum_{j=1}^{2N} a_{j}^\dagger a_{j},
\end{aligned}
\label{H_SSH}
\end{align}
where $a_j^{(\dagger)}$ is a bosonic annihilation (creation) operator for site $j = 1, \dots, 2N$, and $N$ is the number of unit cells.  The two-particle terms, containing the nonlinearity coefficient $v$, create or destroy bosons in \textit{different} sites of a unit cell.  This differs from the more familiar on-site interactions \cite{walls1983, wu1986, gerry2005}, which will be discussed later.

Such off-site nonlinearities can be realized using Josephson parametric converters (JPCs) \cite{bergeal2010}, specifically the three-port JPCs described in Ref.~\onlinecite{Sliwa2015}.  As shown in the bottom panel of Fig.~\ref{fig:f1}(a), each JPC supports three orthogonal modes; by coherently driving one of the three modes, we can establish the desired effective nonlinearity between the two sites (for details, see the Supplemental Materials \cite{SM}).  All the JPCs in the system can be pumped uniformly.  In this context, the mass $m$ represents the detuning between the parametric driving frequency and the natural frequency of the resonators \cite{gerry2005, xi2021}.  A similar mechanism has previously been used in a theoretical proposal for tunable topological phases in quantum photonic lattices \cite{peano2016NC}.  Other photonic platforms with second order nonlinearities, such as microring resonators made of lithium niobate \cite{Tang2022}, may offer alternative approaches for realizing this model.

Two additional sites, representing input and output ports, are now attached to sites $j=1$ and $j=2N$ of the chain, with hoppings $\kappa_{\mathrm{in}}$ and $\kappa_{\mathrm{out}}$.  According to input-output theory \cite{Clerk2010}, the boson operators obey
\begin{align}
  \dot{a}_j &= -i[a_j,\mathcal{H}] - \Gamma_j \,a_j
  - \delta_{j,1}\sqrt{\kappa_{\mathrm{in}}}\, b_{\mathrm{in}}(t),
  \label{InOutput_1} \\
  \dot{a}_j^\dagger &= -i[a_j^\dagger, \mathcal{H}] - \Gamma_j a_j^\dagger
  - \delta_{j,1}\sqrt{\kappa_{\mathrm{in}}}\, \hat{b}_{\mathrm{in}}^\dagger(t),
  \label{InOutput_2} \\
  \Gamma_j &\equiv \gamma 
  + \frac{1}{2} \big(\delta_{j,1}\kappa_{\mathrm{in}}
  + \delta_{j,2N}\kappa_{\mathrm{out}}\big), \label{gammai}
\end{align}
where $b_{\mathrm{in}}(t)$ is the annihilation operator for the input site, a uniform decay rate $\gamma$ has been applied to all sites, and the additional terms in Eq.~\eqref{gammai} describe the hoppings to the input and output ports.

\begin{figure}
\centering
\includegraphics[width=0.48\textwidth]{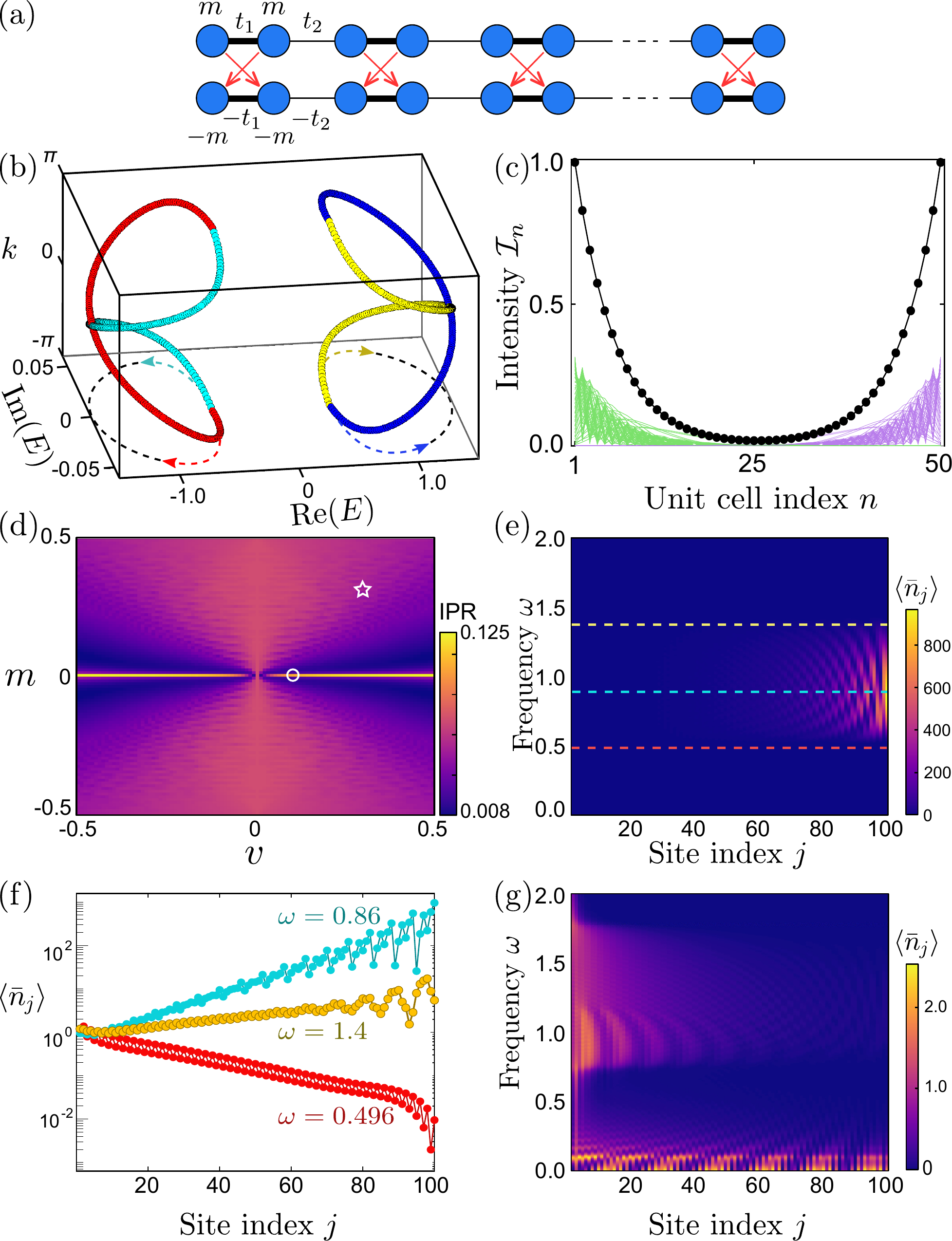}
\caption{(a) Lattice corresponding to the effective Hamiltonian $H_{\mathrm{NH}}$ produced by the BdG transformation, featuring nonreciprocal hoppings $+v$ ($-v$) along (opposite to) the red arrows.  (b) Complex bulk bandstructure of $H_{\mathrm{NH}}$.  The four complex bands for each wavenumber $k$ are drawn in different colors.  Dashes show band projections into the complex $E$ plane, with colored arrows indicating the direction of increasing $k$.  (c) Intensity distribution of the eigenstates of a finite lattice with 50 unit cells and open boundary conditions (OBC).  Green (purple) lines are skin modes on the left (right) boundary, and black dots plot the intensity profile summed over all eigenstates. (d) Heat map of the mean inverse participation ratio (IPR) for eigenstates of the finite lattice with OBC, versus $m$ and $v$.  (e)--(f) Expected mean photon number on each site, $\langle\bar{n}_j\rangle$, at different $\omega$ and $j$, for lattice parameters $m = 0$, $v = 0.1$ [white circle in (d)].  (g) Plot of $\langle\bar{n}_j\rangle$ versus $\omega$ and $j$ for $m = v = 0.3$ [white star in (d)].  Other lattice parameters, used in all subplots, are $t_1=1$, $t_2=0.5$, and $N = 50$.  For (e)--(g), we take $\kappa_{\mathrm{in}}=0.7$, and $\kappa_{\mathrm{out}}=0.1$.}
\label{fig:f3}
\end{figure}

A coherent unit excitation at frequency detuning $\omega$ is applied to the input port, so that the quantum state $|\psi(t)\rangle$ satisfies $b_{\mathrm{in}}(t)\, |\psi(t)\rangle = e^{-i\omega t} |\psi(t)\rangle$ for all $t$.  We can use the Green's function to compute $|\psi(t)\rangle$ \cite{peano2016, mittal2018} (see Supplemental Materials \cite{SM}).  Fig.~\ref{fig:f1}(b) shows the heat map of the average photon number at the output port, $\langle\bar{n}_{\mathrm{out}}\rangle$, versus $\omega$ and the chain length $2N$ (the lattice parameters are stated in the caption). We see that $\langle\bar{n}_{out}\rangle$ increases strongly with $N$ within a wide frequency bandwidth.  Fig.~\ref{fig:f1}(c) shows the line plots at two frequencies, $\omega = 0.86$ (where the highest amplification is achieved) and $\omega = 0$ (away from the amplification band).  In the former case, the graph of $\langle\bar{n}_{\mathrm{out}}\rangle$ versus $N$ is exponential.

Fig.~\ref{fig:f1}(d)--(e) shows the behavior for two other models in which the nonlocal two-particle terms in Eq.~\eqref{H_SSH} are replaced with local terms, $\sum_{j} (v' a^\dagger_j a^\dagger_j + \mathrm{h.c.})$.  In Fig.~\ref{fig:f1}(d), the lattice is not dimerized (i.e., there is one site per unit cell), with inter-site hopping $t_1=1$.  For Fig.~\ref{fig:f1}(e), the lattice is dimerized with intracell (intercell) hoppings $t_1=1$ ($t_2=0.5$).  In each case, the amplification is weaker and less robust.  The highest amplification levels are achieved when $\omega \approx 0$, stemming from lattices modes that are unstable (i.e., having eigenenergies with positive imaginary parts); for details, see the Supplemental Materials \cite{SM}.  We do not observe the exponential increase of $\langle\bar{n}_{\mathrm{out}}\rangle$  with lattice size found in Fig.~\ref{fig:f1}(c).

\section{Characterizing the skin effect}

To understand why the lattice in Fig.~\ref{fig:f1}(a) achieves strong amplification, we use the BdG method to analyze the time evolution of the quadratic bosonic Hamiltonian \cite{Rossignoli2005}.  The Heisenberg equation of motion is
\begin{equation}
  i\frac {\partial{\Psi}}{\partial t}
  = \left[\Psi, \mathcal{ H}\right] = H_{\mathrm{NH}} \Psi,
\label{Evolution}
\end{equation}
where ${\Psi}=({a}_1, \dots, {a}_{2N}, {a}_1^{\dagger}, \dots, {a}_{2N}^\dagger)^T$ and $H_{\mathrm{NH}}$ is a matrix of size $4N\times4N$, $N$ is the total number of unit cells.  Typically, $H_{\mathrm{NH}}$ is non-Hermitian, and it is customary to re-express it using a Hermitian BdG Hamiltonian
\begin{equation}
  H_{\textrm{BdG}} = \tau_z H_{\mathrm{NH}},
\end{equation}
where $\tau_z=\sigma_z\otimes I_{2N}$, $\sigma_z$ is the third Pauli matrix, and $I_{2N}$ is a $2N\times2N$ identity matrix.  But we can opt to analyze the dynamics using $H_{\mathrm{NH}}$.  For instance, if $H_{\mathrm{NH}}$ has eigenvalues with non-zero imaginary parts, Eq.~\eqref{Evolution} would imply that the system is unstable \cite{Rossignoli2005, barnett2013}.  In our case, $H_{\mathrm{NH}}$ describes two coupled chains, with four sites per unit cell and on-site masses and hoppings of opposite signs, as shown in Fig.~\ref{fig:f3}(a).  Crucially, the two-boson terms in $\mathcal{H}$ map to asymmetric inter-chain hoppings $\pm v$, so $H_{\mathrm{NH}}$ is non-Hermitian and nonreciprocal.  For details, see the Supplemental Materials \cite{SM}.

Recent advances in non-Hermitian band theory \cite{Gong2018, Lee2019Anatomy, Okuma2020, Zhang2020} can aid our analysis of $H_{\mathrm{NH}}$.  We first consider the case of $m = 0$.  Fig.~\ref{fig:f3}(b) shows the complex bulk eigenenergies $E$, obtained by applying Bloch periodic boundary conditions (PBC) to one unit cell of the auxiliary lattice with the exemplary parameters $t_1=1$, $t_2=0.5$, and $v=0.1$.  As the Bloch wavenumber $k$ sweeps through $[-\pi,\pi]$, the four bands form two degenerate loops in the complex $E$ plane, each consisting of two bands.  These point gaps imply that the lattice exhibits the NHSE, with the winding determining which end the skin modes are localized to \cite{Gong2018, Lee2019Anatomy, Okuma2020, Zhang2020}.  Here, both winding directions occur, so we expect skin modes to exist on both ends.

One complication is that the correspondence between point gap winding and the NHSE was originally established using non-overlapping bands, whereas each loop in Fig.~\ref{fig:f3}(b) is twofold degenerate.  The situation can be understood by analyzing the symmetries of $H_{\mathrm{NH}}$, which, as described in the Supplemental Materials \cite{SM}, imply that if $E(k)$ is an eigenenergy then so are $E^*(-k)$ and $-E(k)$; moreover, if $H_\mathrm{NH}(k)|\phi_k\rangle=E_k |\phi_k\rangle$, then
\begin{equation}
  H_\mathrm{NH}(-k)|\phi_{-k}\rangle=E_k |\phi_{-k}\rangle,
\end{equation}
where
\begin{equation}
  |\phi_{-k}\rangle=(I_2 \otimes \sigma_x)|\phi_{k}\rangle.
\end{equation}
Therefore, for each $k$ the four Bloch eigenstates split into mutually orthogonal pairs \cite{xue2020}.  For example, in Fig.~\ref{fig:f3}(b) the red and cyan bands are orthogonal, and the yellow and blue bands are orthogonal \cite{SM}.  Each complex $E$ loop is formed by two eigenenergies with opposite windings, whose eigenstates are orthogonal for each $k$.  Such behavior is similar to the $\mathbb{Z}_2$ NHSE described by Okuma \textit{et al.}, where an anti-unitary time reversal symmetry (different from the symmetries described above) gives rise to skin modes that form orthogonal Kramers doublets \cite{Okuma2020}.

Next, we compute the eigenstates and eigenenergies for a finite chain ($N = 50$) under open boundary conditions (OBC).  The eigenenergies are all real, and form arcs enclosed by the eigenenergy loops of the infinite system [Fig.~\ref{fig:f3}(b)], consistent with other models exhibiting the NHSE \cite{Gong2018, Lee2019Anatomy, Okuma2020, Zhang2020}.  All the OBC eigenstates are strongly localized, except for a few with eigenenergies very close to the bulk bands.  Fig.~\ref{fig:f3}(c) shows the intensity profiles for the eigenstates localized on the left (green) and right (magenta), as well as the intensity profile summed over all eigenstates.  (The intensity profile in unit cell $n$ is $\mathcal{I}_{n} = |\psi_{2n-1}|^2 + |\psi_{2n}|^2$, where $\psi_j$ is component $j$ of an eigenvector of $H_{\mathrm{NH}}$; the individual eigenstates are normalized to unity, $\sum_n \mathcal{I}_n = 1$, while the summed intensity profile is normalized to its maximum value.)  There is a macroscopic number of skin modes, which form orthogonal pairs localized on opposite ends of the lattice. If $m \ne 0$, the orthogonality of the bands in each pair is broken.  While $H_{\mathrm{NH}}$ is non-Hermitian, the NHSE is absent \cite{SM}.

Fig.~\ref{fig:f3}(d) shows the mean inverse participation ratio (IPR) \cite{thouless1974} of the eigenstates of $H_{\mathrm{NH}}$ for different $m$ and $v$, for an open chain of $N=50$ unit cells.  We pick out two representative cases.  First, for $m = 0$ and $v = 0.1$, which is the same as in Fig.~\ref{fig:f1} and marked by a circle in Fig.~\ref{fig:f3}(d), $H_{\mathrm{NH}}$ exhibits the NHSE.  For this case, the spatial and frequency distributions of $\langle\bar{n}_j\rangle$, the mean expected photon number on site $j$, are plotted in Fig.~\ref{fig:f3}(e)--(f).  Over a frequency window matching the eigenenergies of the open chain, $\langle\bar{n}_j\rangle$ grows exponentially with $j$.  Outside this band, it decays exponentially with $j$.

This behavior is reminiscent of NHSE-induced amplification in classical or single-particle models \cite{zhang2021acoustic}, but the present analysis applies to a quantum nonlinear multi-boson model with the NHSE entering directly via the BdG transformation.  As all the BdG eigenenergies of the finite chain are real, the amplification comes from the existence of skin modes and not from any dynamical instability, unlike previously-studied models of quantum amplifiers \cite{barnett2013, peano2016, roy2021}. We also observe amplification if the input and output ports are interchanged, as there is another set of skin modes localized to the left. This behavior differs from the unidirectional amplification discussed in Ref.~\onlinecite{McDonald2018}, which arises from non-reciprocal inter-site couplings or non-uniform parametric driving.

\begin{figure}
\centering
\includegraphics[width=0.49\textwidth]{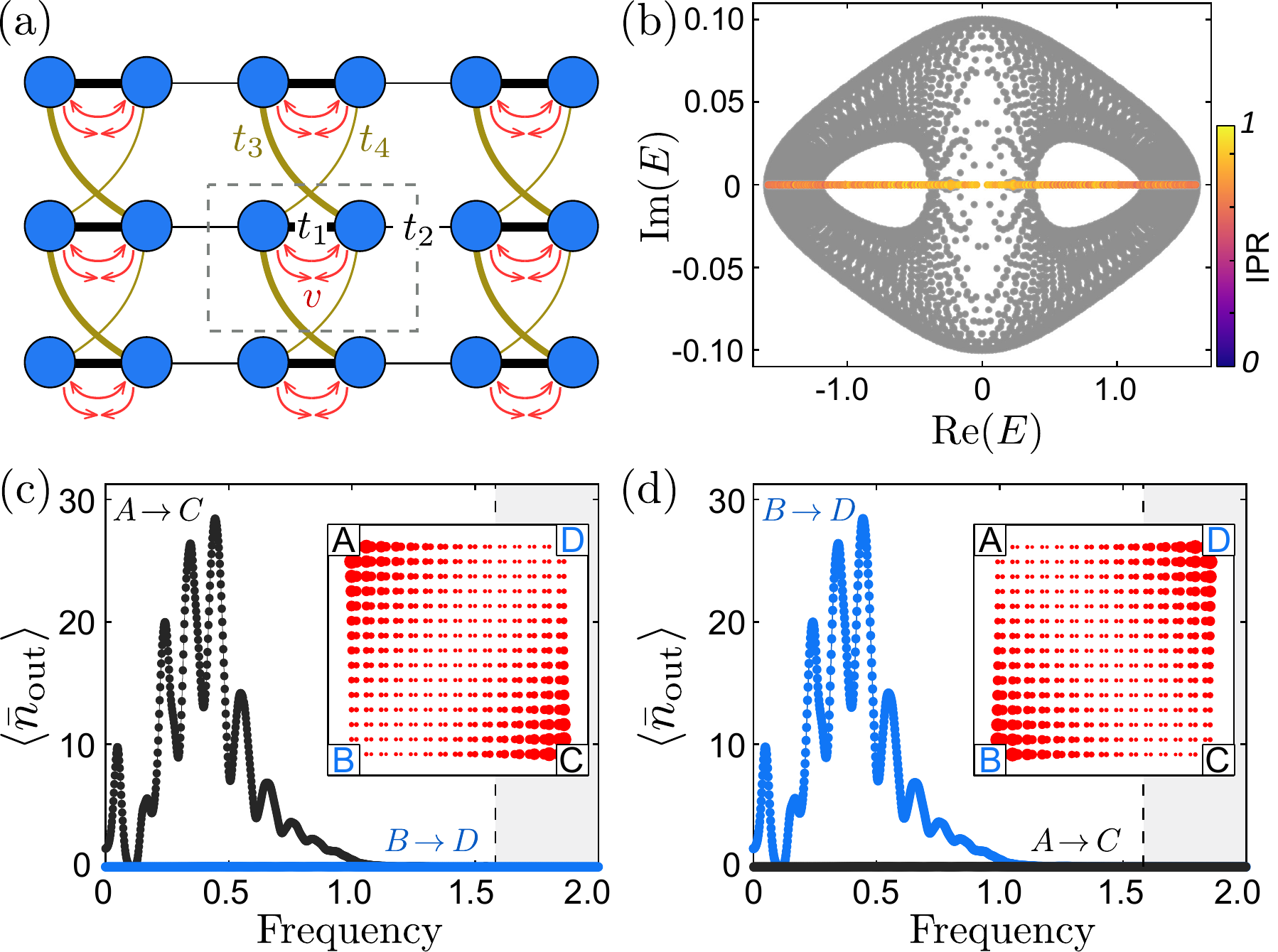}
\caption{NHSE-aided quantum amplification in a 2D lattice. (a) Schematic of the nonlinear 2D lattice, consisting of stacked chains with four atoms (blue circles) per unit cell (dashes).  The thick/thin yellow lines denote hoppings $t_3$ and $t_4$ respectively.  The other features are the same as in the 1D chain in Fig.~\ref{fig:f1}(a).  (b)  Eigenenergies for a lattice of $15\times 15$ unit cells with OBC (colored dots, with colors corresponding to the IPR of each eigenstate), and with PBC (grey dots).  We take $m = 0$, $t_3=0.5$, and $t_4=0.1$, with all other parameters the same as in Fig.~\ref{fig:f1}.  (c) Mean mean photon number at the output port, $\langle\bar{n}_{\mathrm{out}}\rangle$, with input/output ports at different corners of the $15\times 15$ lattice: $A\rightarrow C$ (black dots) and $B\rightarrow D$ (blue dots).  The parameters are the same as in (b).  Inset: summed intensity profile (red dots) for the 2D lattice, with the four lattice corners labeled.  The amplification for $A\rightarrow C$ is consistent with the 2D skin modes occurring at corners $A$ and $C$.  (d) Results after swapping the interchain hoppings, i.e.~$t_3=0.1$, and $t_4=0.5$, with all other parameters unchanged.  Amplification is now observed for $B\rightarrow D$, consistent with the skin modes occurring  at $B$ and $D$ (inset).}
\label{fig:f5}
\end{figure}

Next, we study the case of $m = 0.3$ and $v = 0.3$, marked by a star in Fig.~\ref{fig:f3}(d).  Here, the NHSE is absent; the spectrum under PBC does not form a loop, and overlaps with the spectrum under OBC (see Supplemental Materials \cite{SM}).  In Fig.~\ref{fig:f3}(g), we plot $\langle\bar{n}_j\rangle$ versus $\omega$ and $j$ for the finite chain with the same input/output conditions as before.  Over the entire plotted range, the maximum value of $\langle\bar{n}_j\rangle$ is around 2, compared to around $10^3$ in the amplifying chain [Fig.~\ref{fig:f3}(f)].  There is no exponential spatial amplification along the chain.\\

\section{Two-dimensional lattices}

The NHSE can also induce quantum amplification in higher-dimensional lattices.  Fig.~\ref{fig:f5}(a) shows a 2D lattice formed by stacking copies of the 1D chain of Fig.~\ref{fig:f1}(a), with reciprocal inter-chain hoppings $t_3$ and $t_4$.  Under the BdG transformation, this maps to a single-particle 2D lattice that exhibits a corner NHSE, with all eigenstates localized at certain corners \cite{Lee2019, Zhang2021Observation, zou2021} (see Supplemental Materials \cite{SM}).

In Fig.~\ref{fig:f5}(b), the spectrum under PBC is plotted using gray dots; it occupies a nonzero spectral area, a feature associated with the 2D NHSE \cite{zhang2021}.  The eigenenergies for a $15\times15$ lattice with OBC are plotted with colored dots; they are all real, and the corresponding IPRs are all close to unity, signifying strong localization.

Next, we attach input and output ports to opposite corners of the 2D square lattice.  In Fig.~\ref{fig:f5}(d), we plot $\langle\bar{n}_{\mathrm{out}}\rangle$ versus frequency for the configuration $A\rightarrow C$ (i.e., input at corner $A$ and output at corner $C$, as depicted in the inset), which exhibits a $\sim 30$ fold amplification.  By comparison, for the configuration $B\rightarrow D$ (blue dots), we find $\langle\bar{n}_{\mathrm{out}}\rangle \lesssim 3\times10^{-2}$.  Other configurations such as $A\rightarrow B$ also yield negligible amplification.  This is because the skin modes are predominantly distributed at $A$ and $C$ for these parameters, as indicated by the red circles in the inset, which show the summed intensity profiles for the eigenstates of $H_{\mathrm{NH}}$.  If we swap $t_3$ and $t_4$, the skin modes become localized at $B$ and $D$, and amplification only takes place between those corners, as shown in Fig.~\ref{fig:f5}(d). The 2D NHSE thus enables \textit{directional} quantum amplification.

\section{Conclusion}

We have proposed a way to construct quantum amplifiers based on 1D and 2D arrays of coupled parametrically driven resonators, using the interplay between the non-Hermitian skin effect (NHSE) \cite{Hatano1996, Hatano1997, Yao2018} and the Bogoliubov-de Gennes (BdG) transformation \cite{Rossignoli2005, McDonald2018, Xu2021, Yokomizo2021, Okuma2022}.  Similar effects have also recently been studied in semiclassical exciton-polariton systems \cite{Xu2021}.  The lattices we have introduced can be implemented using three-port Josephson parametric convertors \cite{Sliwa2015}, and can manifest the NHSE without underlying nonreciprocal couplings, and under spatially uniform parametric driving.  The skin modes allow the nonlinear lattice to act as a set of amplifiers in series, without the use of isolators: in a 1D chain, the mean photon number scales exponentially with lattice size, and the amplification occurs over a finite frequency range (corresponding to the bandwidth of the BdG modes), reducing the need for fine-tuning.  Moreover, a 2D lattice can exhibit directional amplification between different points.  In the future, it would be interesting to explore using similar resonator arrays to implement more general classes of nonreciprocal Hamiltonians, which could be useful for manipulating quantum fluctuations.

This work was supported by the Singapore MOE Academic Research Fund Tier~3 Grant MOE2016-T3-1-006 and Tier~1 Grant RG148/20, and by the National Research Foundation Competitive Research Programs NRF-CRP23-2019-0005 and NRF-CRP23-2019-0007.

\bibliography{citepaper.bib}

\begin{widetext}
\newpage

\makeatletter 
\renewcommand{\theequation}{S\arabic{equation}}
\makeatother
\setcounter{equation}{0}

\renewcommand{\thesection}{S\arabic{section}} 
\setcounter{section}{0}

\makeatletter 
\renewcommand{\thefigure}{S\@arabic\c@figure}
\makeatother
\setcounter{figure}{0}

\begin{center}
  \textbf{Supplemental Materials for}\\
  \vskip 0.1in
  {\large ``Amplification of quantum signals by the non-Hermitian skin effect''}\\
  \vskip 0.1in
  {\small Q.~Wang, C.~Y.~Z, Y.~Wang, B.~Zhang, and Y.~D.~Chong}
\end{center}

\section{Input-Output Framework}
\label{sec:input-output}

A quantum signal passing through a complex quantum circuit (consisting of some combination of waveguides, beamsplitters, resonators, etc.)~can be modeled by evolution equations of the form \cite{Clerk2010}
\begin{equation}
\begin{aligned}
  \frac{d{a}_j}{dt} &= -i[{a}_j,{H}]-\gamma{a}_j
  - \left(\kappa_{j,\mathrm{in}}
  + \kappa_{j,\mathrm{out}}\right) {a}_j/2
  - \sqrt{\kappa_{j,\mathrm{in}}}\, a_{\mathrm{in}}(t) \\
  \frac{d{a}_j^\dagger}{dt} &= -i[{a}_j^\dagger,{H}]-\gamma{a}_j^\dagger
  -\left(\kappa_{j,\mathrm{in}} + \kappa_{j,\mathrm{out}}\right)
  {a}_j^\dagger/2
  - \sqrt{\kappa_{j,\mathrm{in}}}\, a_{\mathrm{in}}^\dagger(t),
\end{aligned}
\label{I/O-put-1}
\end{equation}
where ${a}_j$ and ${a}_j^\dagger$ are the time-dependent annihilation and creation operators at site $j$, where $j = 1, \dots, N$; $a_{\mathrm{in}}(t)$ and $a_{\mathrm{in}}^\dagger(t)$ are the annihilation and creation operators for an input signal; $\gamma$ is a background loss on each site; and $\kappa_{j,\mathrm{in}}$ and $\kappa_{j,\mathrm{out}}$ are the couplings from each site of the main lattice to the input and output ports.

By defining  $ {\Psi}=({a}_1,{a}_2,...,{a}_N, {a}_1^{\dagger},{a}_2^{\dagger},...,{a}_N^\dagger)^T$, we can rewrite Eq.~\eqref{I/O-put-1} in the matrix form
\begin{equation}
  \frac{d{\Psi}}{dt} =
  \left(-i H_{NH} - \gamma {I} - C \right) {\Psi} - F,
\label{I/O-put}
\end{equation}
where ${I}$ is the identity matrix, $H_{NH}\Psi=[\Psi, H]$, and the only nonvanishing components of $C$ and $F$ are
\begin{align}
  \left.\begin{aligned}
    C_{jj} &= C_{j+N,j+N}
    = \left(\kappa_{j,\mathrm{in}} + \kappa_{j,\mathrm{out}}\right)/2 \\
    F_j &= \sqrt{\kappa_{j,\mathrm{in}}} a_{\mathrm{in}}(t) \\
    F_{j+N} &= \sqrt{\kappa_{j,\mathrm{in}}} a_{\mathrm{in}}^\dagger(t)
  \end{aligned}\right\}
  \qquad \mathrm{for}~j = 1, \dots, N.
\end{align}
Suppose the input signal has the form
\begin{align}
  a_{\mathrm{in}}(t) = e^{-i\omega t} \; a_{\mathrm{in}}, \;\;\;
  a_{\mathrm{in}}^\dagger(t) = e^{+i\omega t} \; a_{\mathrm{in}}^\dagger.
\end{align}
We take the ansatz
\begin{equation}
  \Psi(t) = \Psi_+ e^{-i\omega t} + \Psi_- e^{i\omega t}.
\label{I/O-put-4}
\end{equation}
Plugging this into Eq.~\eqref{I/O-put} yields
\begin{align}
\begin{aligned}
  \Psi_+ &=
  \left[i(+\omega - H_{NH}) - \gamma {I} - C\right]^{-1} F_+ \,
  a_{\mathrm{in}} \\
  \Psi_- &=
  \left[i(-\omega - H_{NH}) - \gamma {I} - C\right]^{-1} F_- \,
  a_{\mathrm{in}}^\dagger,
\end{aligned}
\label{psisol}
\end{align}
where
\begin{equation}
  \begin{aligned}
  F_+ &= \Big( \sqrt{\kappa_{1,\mathrm{in}}}, \dots,
  \sqrt{\kappa_{N,\mathrm{in}}}, 0, \dots, 0\Big)^T \\
  F_- &= \Big(0, \dots, 0, \sqrt{\kappa_{1,\mathrm{in}}}, \dots,
  \sqrt{\kappa_{N,\mathrm{in}}}\Big)^T.
  \end{aligned}
\end{equation}
Hence, we arrive at a solution of the form
\begin{equation}
  \begin{aligned}
  {a}_j(t) &= p_{j} \, a_{\mathrm{in}} \, e^{-i\omega t}
  + q_{j} \, a_{\mathrm{in}}^\dagger \, e^{+i\omega t} \\
  {a}_j^\dagger(t) &= r_{j} \, a_{\mathrm{in}} \, e^{-i\omega t}
  + s_{j} \, a_{\mathrm{in}}^\dagger \, e^{+i\omega t}
  \end{aligned}
\end{equation}
where the coefficients $p_{j}, q_j, r_j, s_j$ can be extracted from Eq.~\eqref{psisol}, and obey the constraints
\begin{equation}
  r_j = q_j^*, \;\;\; s_j = p_j^*.
  \label{cconstraints}
\end{equation}
At each site $j$, the photon number operator is ${n}_j(t) = {a}_j^\dagger(t) {a}_j(t)$.  We assume the input is a coherent state, so ${b}_{\mathrm{in}}|\alpha\rangle=\alpha|\alpha\rangle$ for some $\alpha \in \mathbb{C}$, with $[{b}_{\mathrm{in}},{b}_{\mathrm{in}}^\dagger ] = 1$.  Then the expectation value of the photon number is
\begin{equation}
  \langle n_j(t) \rangle
  = 2 \mathrm{Re}
  \left[p_j \, q_j^* \, e^{-2i\omega t}\alpha^2\right]
  + \left(|p_j|^2 + |q_j|^2 \right)\, |\alpha|^2
  + \left|q_j\right|^2.
  \label{Photonnumber}
\end{equation}
Taking the time average yields
\begin{equation}
  \langle \overline{n}_j\rangle =
  \left( |p_j|^2 + |q_j|^2\right) \, |\alpha|^2
  + \left|q_j\right|^2.
\end{equation}

\section{Lattice model}
\label{sec:bdgtransform}

Consider the two-site model shown in Fig.~\ref{fig:sf1}(a), which is a one unit cell wide version of the 1D chain discussed in the main text.  Letting $a$ and $b$ denote the bosonic annihilation operators for the two sites, the multi-particle Hamiltonian is
\begin{equation}
  \mathcal{H} = m a^\dagger a + m b^\dagger b
  + t(a^\dagger b+b^\dagger a)+ v(a^\dagger b^\dagger+a b).
  \label{multiH}
\end{equation}
The inter-site hopping is $t$, and $m$ is the on-site mass.  There are quadratic nonlinearity (two-boson interaction) terms acting between the two sites, with coefficient $v$.  We take $m, t, v \in \mathbb{R}$.
\begin{figure}[b]
  \centering
  \includegraphics[width=0.3\textwidth]{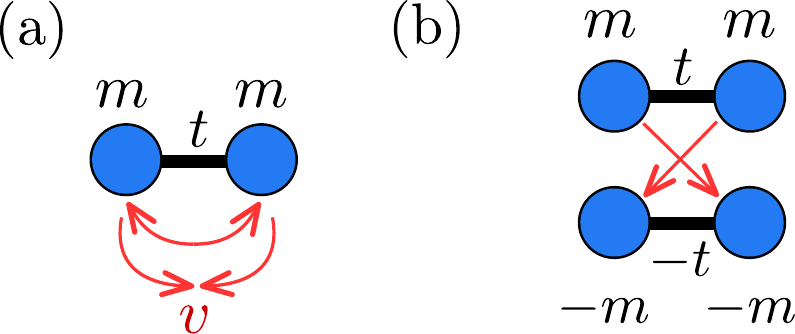}
  \caption{(a) Schematic of a nonlinear two-site model (i.e., a one unit cell wide chain), featuring inter-site hopping $t$ (thick lines), on-site mass $m$, and two-boson creation/annihilation with nonlinearity coefficient $\nu$ (red arrows). (b) The single-particle lattice described by the auxiliary Hamiltonian $H_{\mathrm{NH}}$, which consists of four sites.  The top and bottom sites have opposite signs in their on-site masses and hoppings, and are connected by nonreciprocal hoppings (red arrows).}
  \label{fig:sf1}
\end{figure}

The time-dependent annihilation operators obey the Heisenberg equations of motion
\begin{align}
  i \frac{d a}{d t} &= \left[a(t), \mathcal{H}\right]
  = m \, a(t) + t\, b(t) + v \, b^\dagger(t) \\
  i \frac{d b}{d t} &= \left[b(t), \mathcal{H}\right]
  = m \, b(t) + t\, a(t) + v \, a^\dagger(t),
  \label{heisenberg}
\end{align}
and the creation operators obey the Hermitian conjugates of these equations.  We introduce the Bogoliubov-de Gennes (BdG) transformation by defining
\begin{equation}
  \Psi = \begin{pmatrix}a\\b\\a^{\dagger}\\b^{\dagger} \end{pmatrix}.
  \label{psivector}
\end{equation}
This satisfies the equation of motion
\begin{equation}
  i\frac {\partial{\Psi}}{\partial t}
  = \left[\Psi, \mathcal{ H}\right] = H_{\mathrm{NH}} \Psi,
\label{Evolution}
\end{equation}
where
\begin{equation}
  H_{\mathrm{NH}}=\begin{pmatrix}
             m & t & 0 & v\\
             t & m & v  & 0\\
             0 & -v & -m  & -t\\
             -v & 0 & -t  & -m
             \end{pmatrix}.
\label{H-twosites}
\end{equation}
We refer to $H_{\mathrm{NH}}$ as the BdG Hamiltonian.  It can be interpreted as a single-particle Hamiltonian for the four-site lattice shown in Fig.~\ref{fig:sf4}(b).  For $v \ne 0$, the transformed model has nonreciprocal hoppings, and $H_{\mathrm{NH}}$ is not a symmetric matrix.

Eq.~\eqref{multiH} does not contain the most general form of quadratic nonlinearity; we could also include two-boson interactions that act on the same site.  In Section~\ref{sec:onsite-nonlinearity}, we show that such ``same site'' nonlinearities do not yield the desired non-Hermitian skin effect (NHSE).

This procedure can be generalized to a chain with the multi-particle Hamiltonian
\begin{equation}
  \mathcal{H}=\sum_{n} \left[
    m \left(a_{n}^\dagger a_{n}+ b_{n}^\dagger b_{n}\right)
    + \left(t_1 a_{n}^\dagger b_{n}+t_2 a_{n}^\dagger b_{n-1} +
    v \, a_{n}^\dagger b_{n}^\dagger + \mathrm{h.c.}\right)\right].
\label{H_SSH}
\end{equation}
Here, we use slightly different notation from the main text: $a_n$ and $b_n$ denote the annihilation operators for the two sites in unit cell $n$.  Under periodic boundary conditions (PBC), we can Fourier transform the boson operators to momentum space and apply the BdG transformation:
\begin{equation}
  i\frac{\partial}{\partial t}\begin{pmatrix}a_{k} \\ b_{k}\\ a_{-k}^\dagger\\ b_{-k}^\dagger  \end{pmatrix}
  = H_{\mathrm{NH}}(k)
  \begin{pmatrix}a_{k} \\ b_{k}\\ a_{-k}^\dagger\\ b_{-k}^\dagger  \end{pmatrix}.
\end{equation}
The resulting momentum space BdG Hamiltonian is
\begin{equation}
  H_{\mathrm{NH}}(k)=
  \begin{pmatrix}
    m & t_1+t_2 e^{ik} & 0 & v\\
    t_1+t_2 e^{-ik} & m & v  & 0\\
    0 & -v & -m  & -(t_1+t_2 e^{ik})\\
    -v & 0 & -(t_1+t_2 e^{-ik}) & -m\\
  \end{pmatrix}.
\label{H-SSH-K}
\end{equation}

For $m = 0$ and $v \ne 0$, the 1D chain described by $H_{\mathrm{NH}}$ exhibits the NHSE.  Its complex eigenenergy spectrum is plotted in Fig.~1(c)--(d) of the main text.  Under PBC, the eigenenergies exhibit a point gap in the complex $E$ plane, consisting of two loops.  Each loop is formed by two eigenenergy bands which are orthogonal for the same value of $k$ (hence, a total of four complex bands), and which wind in opposite directions with varying $k$.  Consistent with the previously-established relations between point gap winding and the NHSE \cite{Gong2018, Lee2019Anatomy, Okuma2020, Zhang2020}, the chain under open boundary conditions (OBC) forms skin modes, as shown in Fig.~1(e) of the main text.

\begin{figure}
  \centering
  \includegraphics[width=0.99\textwidth]{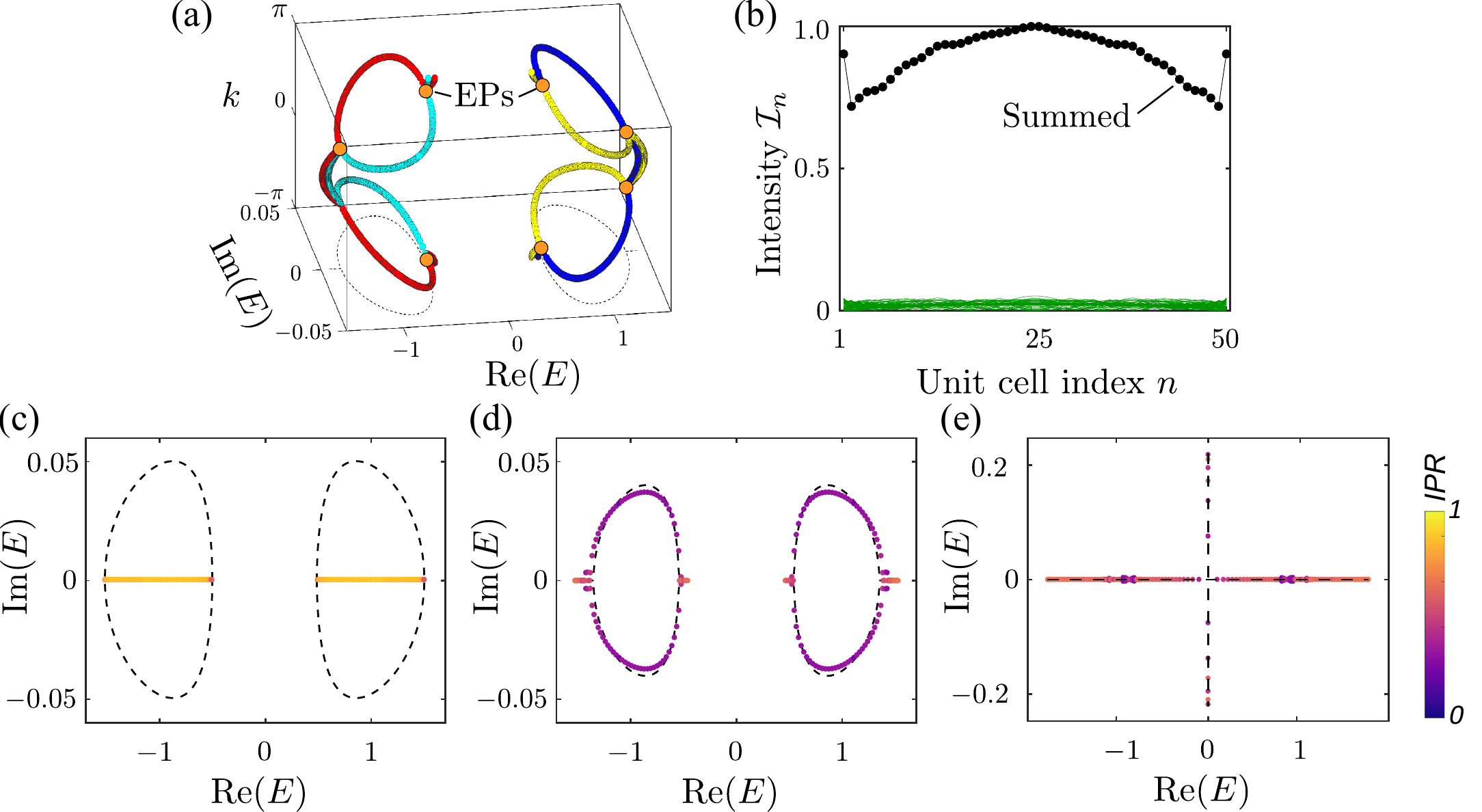}
  \caption{Eigenenergies and eigenmodes of $H_{\mathrm{NH}}$ for different parameters.  (a) Complex bulk bandstructure for $m=0.03$, $t_1=1$, $t_2=0.5$, and $v=0.1$.  For certain values of $k$, the bands coalesce at exceptional points (EPs), indicated by orange circles.  (b) Intensity profiles for the finite lattice with OBC. The green lines show the individual eigenstates, and the black dots are summed over all eigenstates.(c)-(e), Eigenenergies of the finite lattice with 50 unit cells and OBC (colored markers, with colors indicating the inverse participation ratio) and with PBC (black dashes). (c):$m=0$ and $v=0.1$; (d):$m=0.03$ and $v=0.1$; (e):$m=0.3$ and $v=0.3$. For (d) and (e), where NHSEs absent, the OBC and PBC spectra almost overlap.}
  \label{fig:sf2}
\end{figure}

For comparison, Fig.~\ref{fig:sf2} shows the behavior of $H_{\mathrm{NH}}$ when $m = 0.03$, for which the NHSE is not present. The two bands in each pairs are no longer orthogonal to each other, and there are four exceptional points (EPs) \cite{Heiss2012, miri2019} indicated by orange circles in Fig.~\ref{fig:sf2}(a).  At each EP, two eigenstates coalesce (i.e., their eigenvectors become proportional to each other).  The intensity distribution in Fig.~\ref{fig:sf2}(b) shows the absence of the NHSE in this case. In Fig.~\ref{fig:sf2}(c)-(e), we compare the spectrum under OBC and PBC for three different parameter settings: (c) $m=0$ and $v=0.1$; (d) $m=0.03$ and $v=0.1$; and (e) $m=0.3$ and $v=0.3$. For the first case, the NHSE is present and the two spectra are totally different, and the IPR values under OBC are all very large.  For the latter two cases, the OBC and PBC spectra are almost identical, and the IPR values are small indicating the absence of NHSEs. These results also demonstrate that point gaps do not guarantee the existence of the NHSE in multi-band systems.

\section{Symmetries of the Bogoliubov-de Gennes Hamiltonian}

It can be shown that $H_{\mathrm{NH}}(k)$ obeys the following symmetries:
\begin{enumerate}[(I)]
\item Time-reversal symmetry: $TH_{\mathrm{NH}}(k)T=H_{\mathrm{NH}}(-k)$, where $T$ is complex conjugation.

\item $S_0H_{\mathrm{NH}}(k)S_0=-H_{\mathrm{NH}}(k)$, where $S_0 = \sigma_x \otimes I$.

\item $S_xH_{\mathrm{NH}}(k)S_x=H_{\mathrm{NH}}(-k)$, where $S_x=I \otimes \sigma_x$.
\end{enumerate}

\noindent
These have the usual implications for the eigenstates and eigenenergies.  For instance, symmetry (I) implies that if $|\phi\rangle$ is an eigenstate of $H_{\mathrm{NH}}(k)$ with eigenenergy $E$, then $T|\phi\rangle$ is an eigenstate of $H_{\mathrm{NH}}(-k)$ with eigenenergy $E^*$.

Let us now consider the case of $m = 0$, for which $H_{\mathrm{NH}}(k)$ exhibits the NHSE.  As noted in the main text, the skin modes have real eigenenergies, and consist of orthogonal pairs localized to opposite ends of the chain.  This behavior can be understood as follows.  In addition to the symmetries (I)--(III) listed above, $H_{\mathrm{NH}}(k)$ obeys another symmetry if $m = 0$:
\begin{equation}
  GH_{\mathrm{NH}}(k)G=H_{\mathrm{NH}}(k), \;\;\;\mathrm{where}\;\;
  G=\sigma_x \otimes \sigma_z.
  \label{symmetryG}
\end{equation}
This implies that if $H_{\mathrm{NH}}(k)$ has a given eigenstate $(a_k,b_k,c_k,d_k)^T$ with eigenenergy $E_k$, then $(c_k,-d_k,a_k,-b_k)^T$ is an eigenstate of the same Hamiltonian with the same eigenenergy $E_k$.  Away from $k=0$ and $k = \pi$, the eigenvalues are non-degenerate [see Fig.~1 of the main text], so the symmetry must be unbroken.  For a given (complex) band, let
\begin{equation}
  H_{\mathrm{NH}}(k) |\phi_k\rangle = E_k |\phi_k\rangle, \;\;\;\mathrm{where}\;\;
  |\phi_k\rangle = \begin{pmatrix} a_k \\ b_k \\ a_k \\ -b_k \end{pmatrix}.
\end{equation}
Due to symmetry (III),
\begin{equation}
  H_{\mathrm{NH}}(-k) \Big(S_x|\phi_k\rangle\Big) = E_k \Big(S_x|\phi_k\rangle\Big),
  \;\;\;\mathrm{where} \;\;
  S_x|\phi_k\rangle = \begin{pmatrix}b_k\\a_k\\-b_k\\a_k \end{pmatrix}.
  \label{sym3}
\end{equation}
On the other hand, applying symmetry (I) to Eq.~\eqref{sym3} implies that
\begin{equation}
  H_{\mathrm{NH}}(k) |\phi_k'\rangle = E_k^* |\phi_k'\rangle,
  \;\;\;\mathrm{where} \;\;
  |\phi_k'\rangle = TS_x|\phi_k\rangle = \begin{pmatrix} b_k^*\\a_k^*\\-b_k^*\\a_k^* \end{pmatrix}.
\end{equation}
Hence, $H_{\mathrm{NH}}(k)$ has another eigenstate $|\phi_k'\rangle$ with energy $E_k^*$, and the two eigenstates are orthogonal, i.e., $\langle \phi_k' | \phi_{k}\rangle=0$.  For a further discussion of symmetry-induced eigenstate orthogonality in non-Hermitian Hamiltonians, see Ref.~\onlinecite{xue2020}.

Referring to Fig.~2(b) of the main text, this result implies that for each $k$, the red and cyan bands are orthogonal, and the blue and yellow bands are orthogonal.  This structure further implies that the degeneracy points at $k = 0$ and $k = \pi$ are diabolic points (similar to Hermitian systems), rather than exceptional points \cite{xue2020}.  Hence, each of the two eigenenergy loops is formed by a pair of eigenstates that have overlapping trajectories in the complex plane, but are orthogonal to each other.

The above symmetries can aid us in understanding how the NHSE emerges in this system.  First, let us use an intuitive description of the NHSE, based on the formation of standing wave modes, from Ref.~\onlinecite{zhang2021}.  In a 1D lattice with OBC, standing wave eigenstates are formed by superpositions of at least two Bloch waves with the same energy but different $k$ (in order to satisfy the boundary conditions on the two ends).  In a one-band system with a point gap, there is one $k$ for each complex eigenvalue $E$, which obstructs the formation of standing waves \cite{zhang2021}.  In the present system, there are multiple bands forming denegerate loops.  But due to Eq.~\eqref{sym3}, the two eigenstates for any given $E$, which occur at opposite values of $k$, are orthogonal: $\langle\phi_k | S_k |\phi_k\rangle = 0$.  This acts as a special obstruction for using these two eigenstates to simultaneously satisfy the two boundary conditions, leading to the NHSE.  For $m \neq 0$, the symmetry \eqref{symmetryG} does not apply, and hence there is no eigenstate orthogonality and the NHSE does not occur \cite{okuma2019, lu2021}.

\section{Origin of the on-site mass and its effect on the amplifier}

In this section, we will review why the on-site mass $m$ arises in a parametrically driven resonator as the detuning between the driving frequency and the resonant frequency of the cavities \cite{gerry2005, xi2021}.  We will then describe how varying $m$ affects the performance of the quantum amplifier described in the main text.

Consider a two site case with the initial Hamiltonian
\begin{equation}
\mathcal{{H}}=w_0({a}^\dagger {a}+{b}^\dagger {b})+\omega_p {c}^\dagger {c}+(\chi {a}^\dagger {b}^\dagger {c}+\chi^*{a}{b}{c}^\dagger)
\label{H0}
\end{equation}
where $a$ and $b$ are signal modes with frequency $\omega_0$, $c$ is a pump mode with frequency $\omega_p$, $\chi$ is a second order nonlinear cofficient.  We assume the pump field is in a coherent state $|\alpha e^{-i\omega t}\rangle$, and take the approximation
\begin{equation}
  \begin{aligned}
    {c}|\alpha e^{-i\omega_pt} \rangle \approx \alpha e^{-i\omega_pt}, \\
    {c}^\dagger|\alpha e^{i\omega_pt} \rangle \approx \alpha^* e^{i\omega_pt}.
  \end{aligned}
\end{equation}
Thus,
\begin{equation}
\mathcal{{H}}^P=\omega_0({a}^\dagger {a}+{b}^\dagger {b})+(v^* e^{-i\omega_pt} {a}^\dagger {b}^\dagger+ v e^{i\omega_pt} {a}{b}).
\label{Hp}
\end{equation}
Here, we have dropped the constant $\omega_p|\alpha|^2$, and defined $v=\chi^* \alpha^*$. We then define
\begin{equation}
  \begin{aligned}
    W &= \omega_p({a}^\dagger {a}+{b}^\dagger {b})/2 \\
    V(t) &= \mathcal{{H}}^P-W.
  \end{aligned}
\end{equation}
For any operator $A$ (e.g., the creation operator $a^\dagger$), the Heisenberg equation is
\begin{equation}
i \frac{\partial {A(t)}}{\partial t} = \left[A(t),\mathcal{{H}}^P\right].
\label{A1}
\end{equation}
We go to the interaction picture by defining
\begin{equation}
U(t)=e^{i\omega t} \Rightarrow  \left\{
\begin{aligned}
\frac{\mathrm{d} {U(t)}}{\mathrm{d} t} &= iWU(t)\\
\frac{\mathrm{d} {U^\dagger (t)}}{\mathrm{d} t} &= -iU^\dagger (t) W
\end{aligned}
\right.
\label{A2}
\end{equation}
and
\begin{equation}
A_I(t)=U^\dagger(t) A U(t).
\label{A3}
\end{equation}
Then
\begin{equation}
\begin{aligned}
i \frac{\mathrm{d} {A_I}}{\mathrm{d} t}
&=i \frac{\mathrm{d} U^\dagger}{\mathrm{d}t} A U+i U^\dagger  \frac{\mathrm{d} A}{\mathrm{d}t} U+iU^\dagger A \frac{\mathrm{d} U}{\mathrm{d}t}\\
&=i[A_I, H_I]
\end{aligned}
\label{A4}
\end{equation}
with 
\begin{equation}
H_I=U^\dagger [\mathcal{{H}}^P-W] U=(\omega_0-\omega_p/2)({a}^\dagger {a}+{b}^\dagger {b})+(v^*  {a}^\dagger {b}^\dagger+ v {a}{b})
\label{A5}
\end{equation}
We can thus define $m=\omega_0-\omega_p/2$.  Its value can be set by tuning $\omega_0$ or $\omega_p$ \cite{xi2021}.

\begin{figure}
  \centering
  \includegraphics[width=0.8\textwidth]{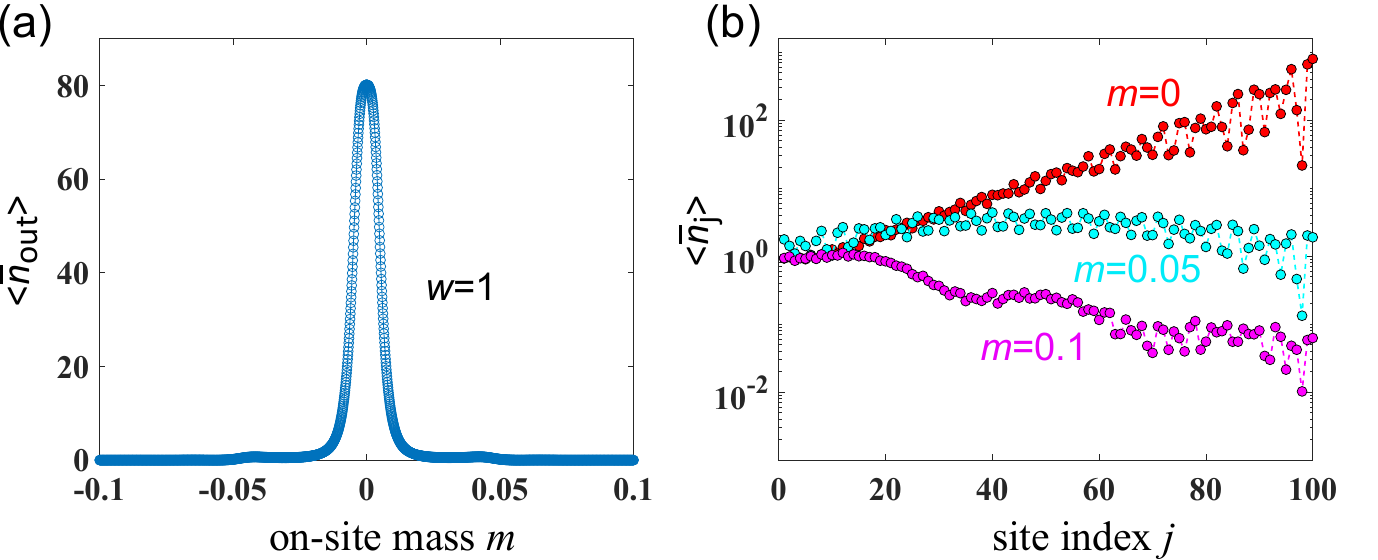}
  \caption{(a), mean photon number from the output port versus $m$ with frequency $w=1$. (b), semi-logarithmic plot of $\langle\bar{n}_j\rangle$ versus $j$ at three frequencies with red, cyan, and magneta dots indicate $m$ take value of 0,  0.05 and 0.1 respectively. The setup is same as Fig. 1(a) in manuscript. Parameters are set $t_1=1$, $t_2=0.5$, $v=0.1$, $\kappa_{\mathrm{in}}=0.7$, $\kappa_O=0.1$ and $\gamma=0.02$. }
  \label{fig:sf3}
\end{figure}

In Fig.~\ref{fig:sf3} (a), we show the mean photon number from the output port versus on-site mass for frequency $w=1$. The setup and other parameters of the lattice are all same as Fig.1 (a) in manuscript, and the result show the amplifier works approxmiately for $m\in[-0.01, 0.01]$. In Fig.~\ref{fig:sf3}(b), we show the distribution of the mean photon number at each site, with red, cyan, and magneta dots indicate $m$ take value of 0,  0.05 and 0.1 respectively.

\section{Lattice Hamiltonian without the NHSE}
\label{sec:onsite-nonlinearity}

\begin{figure}[htb]
\centering
\includegraphics[width=0.66\textwidth]{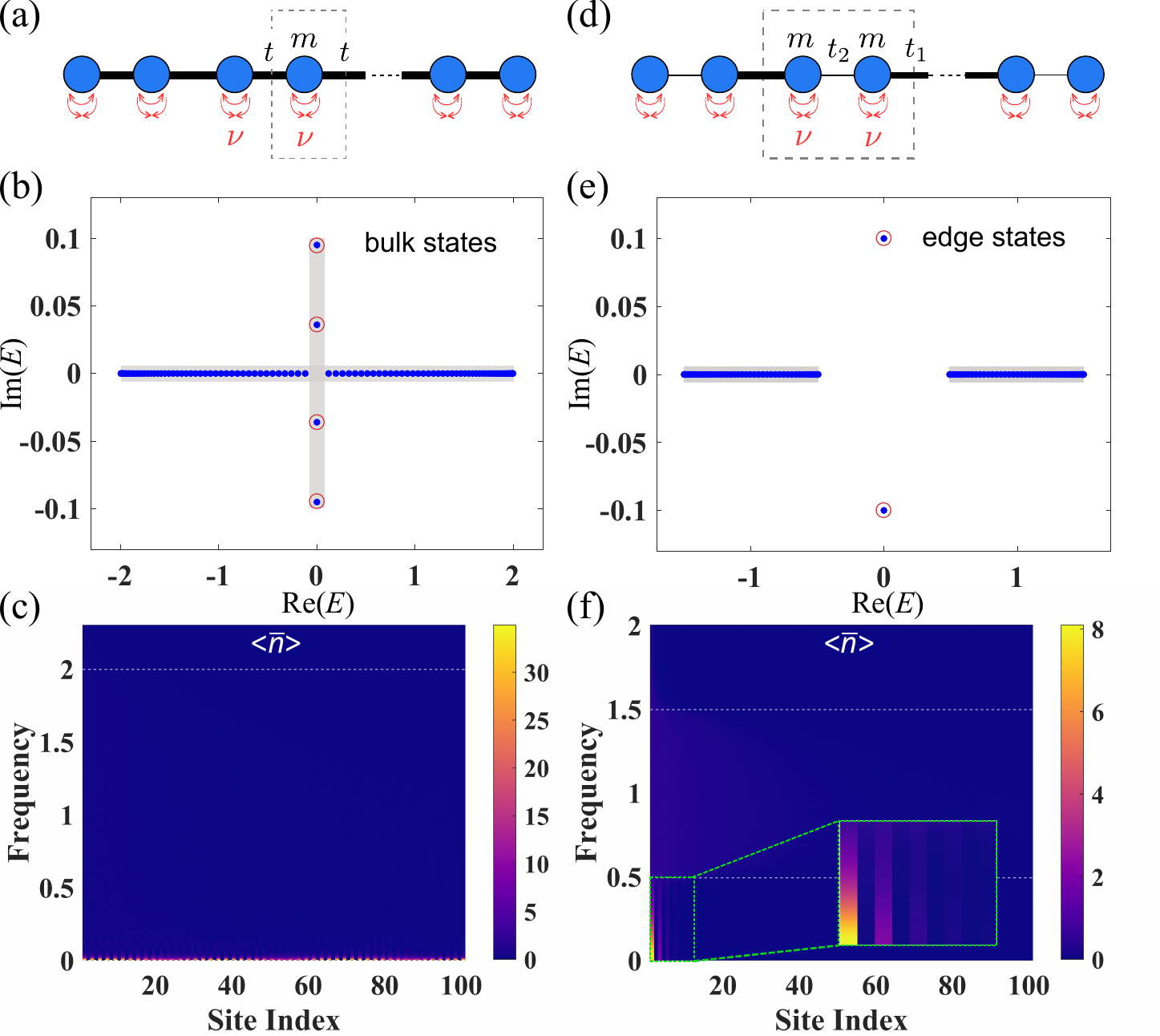}
\caption{Results for lattices without NHSEs. (a), A simple lattice model with one site in each unit cell, the coupling strength between two nearest sites is $t=1$, and the two photon driving term is operated on each sites. (b), Complex eigenenergy spectrum. The eigenenergy under OBC was plot with blue dots, and the bulk states with pure imaginary eigenvalues are highlight by the red circles; the spectrum under PBC are marked by the shick gray lines. (c), Heat map of $\langle\bar{n}_j\rangle$ versus $j$ and $\omega$, showing the amplification only works for frequency near $\omega=0$. (d), A conventional SSH lattice with two photon driving term operated on each same sites, with coupling strength $t_1=1$ and $t_2=0.5$. (e-f), same as (b-c). The other parameters, used in all subplots, are $m=0$, $\kappa_{\mathrm{in}}=0.7$, $\kappa_{\mathrm{out}}=0.1$, $\gamma=0.02$, and the finite lattice consist of $100$ sites. }
\label{fig:sf4}
\end{figure}

The lattice Hamiltonian is
\begin{equation}
  \mathcal{ H}=\sum_{j=1}^N \{ m a_{j}^\dagger a_{j}\}+
                   \sum_{j=1}^N\{t a_{j-1}^\dagger a_{j}+h.c.\}+
                   \sum_{j=1}^N \{v a_{j}^\dagger a_{j}^\dagger+h.c.\} 
\label{H_Single}
\end{equation}
here $j$ labels the site index.  The on-site mass was $m=0$, the coupling between two nearest sites was $t=1$, and the two-particle driving with strength $v=0.05$ was operated on each same sites. The eigenvalues for a finite chain that consists of 100 sites are plot in Fig.~\ref{fig:sf4}(b) with blue dots, and the spectrum for PBC are plot with thick gray lines, which covers the same regime. The unstable bulk (with non-zeros imaginary parts) are highlight by the red circles. The Heat map of $\langle\bar{n}_j\rangle$(the expected time-averaged photon number on site $j$) versus $j$ and $\omega$ with the same setup as Fig. 1(d) in manuscript was shown in Fig.~\ref{fig:sf4}(c).

The lattice as shown in Fig.~\ref{fig:sf4}(d)(Fig. 1(e) in manuscript) was similiar with  conventional SSH model\cite{Su1979}, and the total Hamiltonian can be written as: 
\begin{equation}
\begin{aligned}
  \mathcal{ H}=&\sum_{j=1}^N \{ m a_{2j-1}^\dagger a_{2j-1}+m a_{2j}^\dagger a_{2j}\}+
                   \sum_{j=1}^N\{t_1 a_{2j-1}^\dagger a_{2j}+t_2 a_{2j-1}^\dagger a_{2j-2}+h.c.\}+\\
                   &\sum_{j=1}^N \{v a_{2j}^\dagger a_{2j}^\dagger+v a_{2j-1}^\dagger a_{2j-1}^\dagger+h.c.\} 
\end{aligned}
\label{H_SSH}
\end{equation}
here $j$ labels the unit cell index.  The on-site mass is $m=0$ for both two sites, the intra-cell coupling $t_1=1$ and inter-cell coupling $t_2=0.5$ was indicated by the thick and thin lines, also the two-particle driving with strength $v=0.05$ was operated on each same sites. The eigenvalues for a finite chain that consists of 100 sites are plot in Fig.~\ref{fig:sf4}(e) with blue dots, and the spectrum for PBC are plot with thick gray lines, which covers the same regime. The unstable topological edge modes (with non-zeros imaginary parts) are highlight by the red circles. The Heat map of $\langle\bar{n}_j\rangle$  versus $j$ and $\omega$ with the same setup as Fig. 2(a) in manuscript was shown in Fig.~\ref{fig:sf4}(f), and the insert shows the details near $\omega=0$.

\section{Hamiltonian for the 2D model}

The Hamiltonain for the 2D model discussed in the main text is
\begin{align}
\begin{aligned}
  \mathcal{H} &=
  \sum_{x,y=1}^{N_x,N_y} \Bigg[ \left(t_1 a_{2x-1,y}^\dagger a_{2x,y}+t_2 a_{2x-1,y}^\dagger a_{2x-2,y}+ \mathrm{h.c.}\right)\\
    & \qquad\qquad + \left(t_3 a_{2x-1,y}^\dagger a_{2x,y+1}+t_4 a_{2x,y}^\dagger a_{2x-1,y+1}+ \mathrm{h.c.}\right)\\
    &\qquad\qquad + \left( \nu\, a_{2x-1,y}^\dagger a_{2x,y}^\dagger+ \mathrm{h.c.}\right) \Bigg] \\
  &\qquad\qquad\quad+ m \sum_{x,y=1}^{N_x,N_y}\left(a_{2x-1,y}^\dagger a_{2x-1,y}+a_{2x,y}^\dagger a_{2x,y}\right),
\end{aligned}
\label{H_2D}
\end{align}
where $x$ and $y$ indicate the position index of each unit cell, and $x\in[1, N_x]$ and $y\in[1, N_y]$,  $N_x$ and $N_y$ is the number of unit cells along $X$ and $Y$ axises.  From this, we can get the single-particle Hamiltonian $H_{\mathrm{NH}}$ by following the same procedure as in Sections~\ref{sec:input-output}--\ref{sec:bdgtransform}.

A rigorous understanding of the necessary conditions for the emergence of the NHSE in two- and higher-dimensions is still lacking.  Usually, a non-zero spectral area in momentum space implies the existence of the NHSE \cite{zhang2021}, though this may be dependent on the shape of the finite lattice (i.e., the NHSEs may be absent for some shapes, but present for others). In our model, the PBC spectrum has non-zero area, as shown in Fig.~3(b) of the main text, and the square-shaped finite lattice can indeed host corner NHSEs.

\clearpage

\end{widetext}
\end{document}